\documentclass[11pt,twoside]{article}
\usepackage{asp2010}

\resetcounters

\begin{document}

\title{The Age-Rotation-Activity Relation: From Myrs to Gyrs}
\author{Kevin R.\ Covey$^{1,2}$, Marcel A.\ Ag\"ueros$^3$, Jenna J.\ Lemonias$^3$, Nicholas M.\ Law$^4$, Adam L.\ Kraus$^5$, and the Palomar Transient Factory Collaboration
\affil{$^1$Hubble Fellow; Cornell University, Department of Astronomy, 226 Space Sciences Building, Ithaca, NY 14853, USA}
\affil{$^2$Visiting Researcher, Department of Astronomy, Boston University, 725 Commonwealth Avenue, Boston, MA 02215, USA}
\affil{$^3$Department of Astronomy, Columbia University, 550 West 120th Street, New York, NY 10027, USA}
\affil{$^4$Dunlap Institute for Astronomy and Astrophysics, University of Toronto, 50 St.\ George Street, Toronto M5S 3H4, Ontario, Canada}
\affil{$^5$Hubble Fellow; University of Hawaii-IfA, 2680 Woodlawn Drive, Honolulu, HI 96822, USA}}

\begin{abstract}
Over the past 40 years, observational surveys have established the existence of a tight relationship between a star's age, rotation period, and magnetic activity. The age-rotation-activity relation is essential for understanding the interplay between, and evolution of, a star's angular momentum content and magnetic dynamo. It also provides a valuable age estimator for isolated field stars. While the age-rotation-activity relation has been studied extensively in clusters younger than 500 Myr, its subsequent evolution is less constrained. Empirically measured rotation periods are scarce at intermediate ages (i.e., Hyades or older), complicating attempts to test reports of a break in the age-activity relation near 1 Gyr \citep[e.g.,][]{pace2004,giardino2008}.

Using the Palomar Transient Factory (PTF), we have begun a survey of stellar rotation to map out the late-stage evolution of the age-rotation-activity relation: the Columbia/Cornell/Caltech PTF (CCCP) survey of open clusters. The first CCCP target is the nearby $\sim$600 Myr Hyades-analog Praesepe; we have constructed PTF light curves containing $>$150 measurements spanning more than three months for $\sim$650 cluster members. We measure rotation periods for 40 K \& M cluster members, filling the gap between the periods previously reported for solar-type Hyads \citep{Radick1987,Prosser1995} and for a handful of low-mass Praesepe members \citep[][]{scholz2007}.  

Our measurements indicate that Praesepe's period-color relation undergoes at transition at a characteristic spectral type of $\sim$M1 --- from a well-defined singular relation at higher mass, to a more scattered distribution of both fast and slow-rotators at lower masses.  The location of this transition is broadly consistent with expectations based on observations of younger clusters and the assumption that stellar-spin down is the dominant mechanism influencing angular momentum evolution at $\sim$600 Myr.  Combining these data with archival X-ray observations and H$\alpha$ measurements provides a portrait of the $\sim$600 Myr age-rotation-activity relation (see contribution by Lemonias et al.\ in these proceedings). In addition to presenting the results of our photometric monitoring of Praesepe, we summarize the status and future of the CCCP survey.
\end{abstract}

\section{Introduction}
In his seminal 1972 paper, Andrew Skumanich showed that stellar rotation decreases over time such that $v_{rot} \propto t^{-0.5}$ --- as does CaII emission, a measure of chromospheric activity and proxy for magnetic field strength. This relationship between age, rotation, and activity has been a cornerstone of stellar evolution work over the past 40 years, and has generated almost as many questions as applications. For example, angular momentum loss due to stellar winds is generally thought to be responsible for the \citet{skumanich72} law, but the exact dependence of $v_{rot}$ on age is not entirely clear, and relies on the assumed stellar magnetic field geometry and degree of core-envelope coupling \citep{kawaler1988, krishnamurthi1997}. Furthermore, later-type, fully convective stars appear to have longer active lifetimes than their early-type brethren \citep[e.g.,][]{andy08}, indicating that the lowest mass stars are capable of generating significant magnetic fields even in the absence of a standard solar-type dynamo \citep{Browning2008}.  The lack of a comprehensive theoretical understanding of the age-rotation-activity relation has not prevented the development and use of gyrochronology, however, which attempts to determine the precise ages of field stars based on a presumed age-rotation relation \citep[e.g.,][]{barnes2007, mamajek2008, collier2009, barnes2010}, nor of empirical age-activity relations, which do not always find activity decaying with time quite as simply as predicted by the Skumanich law \citep[e.g.,][]{feigelson04,pace2004,giampapa2006}.

Fully mapping out the dependence of stellar rotation and activity on age requires the study of stars ranging in both mass and age. Statistical constraints on the age-rotation-activity relation can be derived via analysis of Galactic field stars \citep[e.g.,][]{feigelson04,Covey2008,Irwin2010}, but the homogeneous, coeval populations in open clusters provide an ideal environment for studying time-dependent stellar properties. There are relatively few nearby open clusters, however, and fewer still have the high quality optical data needed to characterize their rotations --- in part because of the sheer difficulty involved in systematically monitoring a large number of stars over several months or more. As a result, our current view of the age-rotation-activity relation depends on observations of handfuls of stars in the field and in a small number of well-studied clusters \citep[with the Hyades being a particularly key cluster, e.g.,][]{Radick1987,jones1996,stauffer1997,terndrup2000}. 

The advent of time-domain surveys, with their emphasis on wide-field, automated, high-cadence observing, makes it possible to monitor stellar rotation in clusters on an entirely new scale \citep[e.g.,][]{Irwin2007,Meibom2009,Hartman2010}. Many ongoing time-domain surveys are primarily designed to identify transiting exoplanets, however, and thus aim to cover the widest area possible to a relatively modest depth.  Deep targeted surveys of open clusters are therefore still required to measure rotation periods, particularly for the lowest mass stars in older, more distant open clusters. The Palomar Transient Factory provides deep, multi-epoch photometry over a wide field-of-view, and our Columbia/Cornell/Caltech PTF survey is leveraging this capability to map $v_{rot}(t)$ in open clusters of different ages. Our first CCCP target, Praesepe (the Beehive Cluster, M44, $08^h 40^m 24^s +19^{\circ} 41\arcmin$), is a nearby ($\sim$170 pc), intermediate-age ($\sim$600 Myr), and rich \citep[its membership has recently been expanded to $\sim$1200 stars;][]{adam2007} open cluster that shares many characteristics with the Hyades.

Here we report the stellar rotation periods for Praesepe members derived from our first season of PTF observations. Our campaign produced $\sim$200 distinct observations of four overlapping fields covering a $3.75\times 3.30$ deg area designed to include a large number of Praesepe members identified by \citet{adam2007}. In Section~\ref{observations} we describe our PTF observations. We discuss our period-finding algorithm in Section~\ref{periods} and our results in Section \ref{results}. Finally, in Section~\ref{future} we outline the current status and future of the CCCP survey.

\section{Observations}\label{observations}
PTF is a transient detection system comprised of a wide-field survey camera mounted on the automated Samuel Oschin 48 inch telescope at Palomar Observatory, CA (known as the P48), an automated real time data reduction pipeline, a dedicated photometric follow-up telescope (the Palomar 60 inch), and an archive of all detected sources \citep[for details, see][]{nick2009,rau2009}. The survey camera has 101 megapixels, 1$\arcsec$ sampling and a 8.1 deg$^2$ field-of-view covered by an array of 12 CCD chips, one of which is now inoperative. Currently, observations are performed in one of two filters: SDSS-$g$ or Mould-$R$. Under the median 1.1$\arcsec$ seeing conditions the camera produces 2.0$\arcsec$ FWHM images and reaches $5\sigma$ limiting AB magnitudes of $m_{g} \approx\ 21.3$ and $m_R \approx\ 20.6$ mag in its standard 60 s exposures. 

Praesepe was monitored by PTF for 3.5 months in early 2010, beginning on 2010 Feb 2 and ending on 2010 May 19.  Our observing cadence was sensitive to $P_{rot}$ from a few to a few hundred hours, covering the range occupied by the few cluster members with known periods \citep[][]{scholz2007}. PTF monitored Praesepe by imaging four overlapping 3.5 x 2.31 deg fields, which together cover $\sim$18 deg$^2$ in the cluster's center.  The extent of the CCCP footprint is shown in Fig.~1; $>$80,000 individual objects were detected in our observations. 

\begin{figure}[!ht]
\plotone{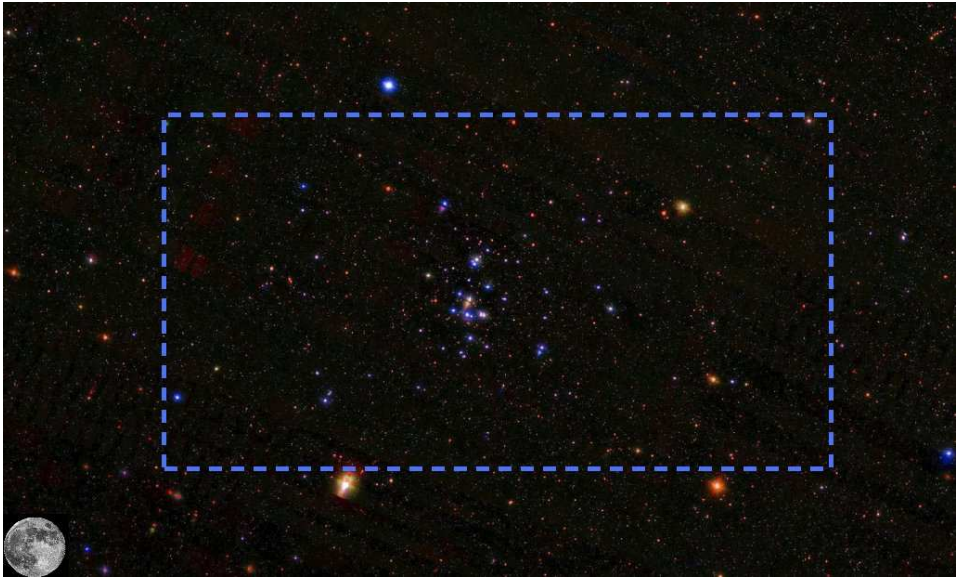}
\caption{SDSS image of the Praesepe open cluster. The $\sim$18 deg$^2$ monitored by the CCCP survey is shown by the blue dashed rectangle; the Moon is shown to scale in the lower left corner of the image.}
\end{figure}

In developing their cluster catalog, \citet{adam2007} combined archival data from multiple surveys to calculate proper motions and photometry for several million sources within 7 deg of Praesepe's center. Their census covers a larger area of sky and is deeper than any previous proper motion study, and extends to near the stellar/sub-stellar boundary. The resulting catalog includes 1169 candidate members with membership probability $>$50\% (hereafter referred to as the P50 stars); 442 are identified as high-probability candidates for the first time. \citet{adam2007} estimate that their survey is $>$90\% complete across a wide range of spectral types, from F0 to M5.

Of the 1169 P50 members in the \citet{adam2007} catalog, 923, or close to 80\%, lie within the CCCP footprint.  Of these, 661 are fainter than the PTF saturation limit ($\sim$14 mag): PTF detected 534 (or 81\%) of these candidate members, with the rest falling within chip gaps or on the dead chip. \citet{adam2007} provide spectral types for Praesepe stars based on spectral energy distribution fitting; the PTF-detected members are late K through early M stars, which is as expected given the distance to Praesepe and the PTF exposure time.  Aperture photometry was measured for each candidate member at each
epoch using the SExtractor software \citep{Bertin1996}. Positional
matching was then used to merge detections across epochs,
producing a single light curve for sources. The photometric zeropoints
were adjusted on a per chip basis to minimize the median photometric
variability of the detected sources. For more information on this
reduction process, see Law et al. 2010 elsewhere in these
proceedings

\section{Period Measurements}\label{periods}
We use a modified version of the Lomb-Scargle algorithm to search our light curves for periodic signals due to rotation.  The properties of each star's light curve defined the range of frequencies that were searched for periodic signals: we followed Eq.\ 11 of \citet{frescura2008} in calculating the frequency grid from the number of individual measurements obtained for each star, as well as the total duration of the light curve.  We oversampled this grid by a factor of five to ensure maximum sensitivity to any periodic behavior sampled by our monitoring.   Potential beat frequencies between the primary periodogram peak and a possible one-day alias, typical for ground-based, nightly observing campaigns, were flagged following Eq.\ 1 of \citet{messina2010}.  

To test the significance of the periods identified by our modified Lomb-Scargle algorithm, we performed a Monte Carlo analysis of our light curves \citep[for a similar analysis, see][]{sturrock2010}. Having measured potential rotation periods for all cluster members detected by PTF, we then conducted an identical analysis on each light curve after randomly scrambling the magnitudes measured at each epoch.  Repeating this test 100 times on each scrambled light curve, we were able to identify the maximum measured periodogram peak as the power threshold corresponding to a $<$1\% false alarm probability (FAP) in the absence of ordered variations.  This analysis established that, across our entire sample, a periodogram peak with power $>$25 corresponded to a FAP $<$1\%; indeed, for only three of the 534 stars analyzed here did the 1\% FAP correspond to a periodogram power threshold $>$20.  

Adopting a conservative power threshold of 30 to select potentially periodic cluster members, we then visually inspected the output of our search for each candidate.  Periodograms were checked to confirm the presence of a single narrow peak, well separated from the underlying background power; further scrutiny established that the periodic behavior was visible and stable across the full light curve, well sampled in phase, and of an amplitude at least comparable to the observational noise.  Representative periodograms and phased light curves for the resulting high-confidence detections are shown in Fig. 2.

\begin{figure}[!ht]
\plotone{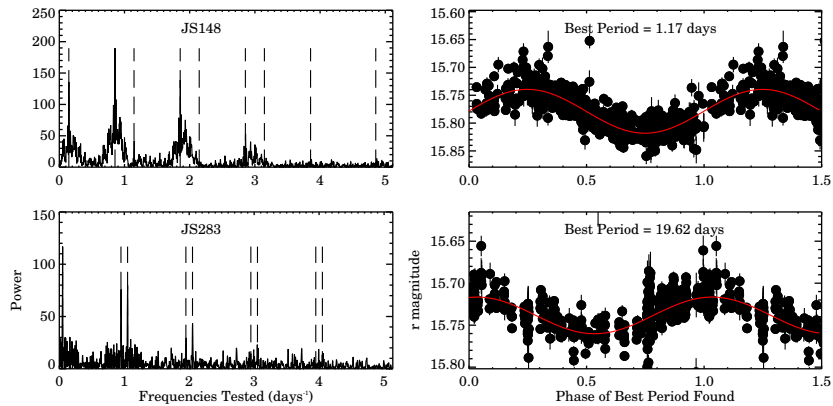}\label{example}
\caption{Example periodograms (left panels) and phased light curves (right panels) for two P50 Praesepe stars with high-confidence rotation period measurements. A fast and slow rotator are shown in the top and bottom panels, respectively.}
\end{figure}

\section{Results}\label{results}
\begin{figure}[!ht]
\plotone{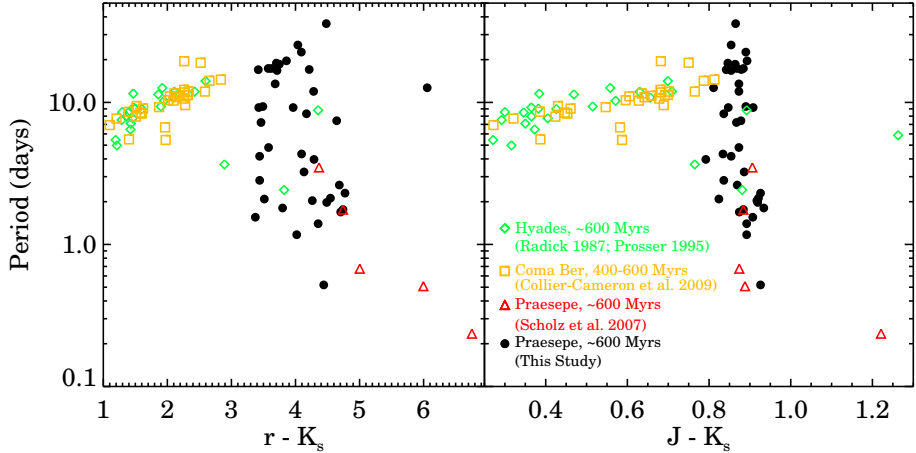}\label{mp}
\caption{Color-period relations for stars in moderately old (t$\sim$600 Myr) open clusters.  Stellar periods are plotted as a function of SDSS $r$ - 2MASS K$_s$ color (left panel) and 2MASS J-K$_s$ color (right panel).  All stars possess native 2MASS J-K$_s$ colors.  $r$-K$_s$ colors were constructed for Praesepe stars from native SDSS $r$ and 2MASS K$_s$ magnitudes.  For Hyades stars, $r$-K$_s$ colors were determined using synthetic $r$ magnitudes calculated from observed B and V magnitudes \citep{Joner2006} and the $r$(B,V) transformation defined by \citet{Jester2005}.  For Coma Ber stars, synthetic $r$-K$_s$ colors were estimated from observed J-K$_s$ colors using the median SDSS-2MASS stellar locus defined by \citet{Covey2007}. }
\end{figure}

The analysis described above produces high-confidence measurements of rotation periods ranging from 0.45 to 35.85 days for a total of 40 stars. The CCCP data clearly indicate the presence of both fast and slow rotators in Praesepe; examples of a fast and slow rotator are shown in Fig.~\ref{example}, and the location of the full sample in color/period space is shown in Fig.\ref{mp}, along with additional stars from comparably aged clusters (i.e., Hyades \& Coma Ber).  These observations, along with complementary measurements of low-mass Hyades members (see contribution by Delorme et al.\ elsewhere in this proceedings), establish that the $\sim$600 Myr period-color relation is single-valued for colors blueward of $r-$K$_s \sim 3$ and J-K$_s \sim 0.825$, corresponding to a spectral type of $\sim$M1 and a mass of M $\sim$ 0.6 M$_{\odot}$ \citep{Bochanski2007,adam2007}.  Redward of this color, however, the distribution of stars in color-period space is scattered, with populations of both fast and slow rotators.  

The morphology of the color-period relation we measure in Praesepe is broadly consistent with observations of somewhat younger clusters, assuming that stellar spin-down is the dominant mechanism governing the stellar angular momentum at these ages.  The color-period relation observed in the slightly younger M37, for example, departs from a single-valued relation at M $\sim$ 0.8 M$_{\odot}$ \citep{hartman2009}, suggesting that stars with 0.8 M$_{\odot} < $ M $<$ 0.6 M$_{\odot}$ possess spin-down timescales of $\sim$500-650 Myr.  To go beyond this crude comparison, we are currently analyzing the color-period relation we measure in Praesepe in light of the gyrochronology relations presented in the literature \citep[e.g.,][]{mamajek2008,barnes2010}.  

This spin-down timescale also agrees well with the $\sim$600 Myr magnetic activity lifetime predicted for $\sim$M1 stars based on statistical analyses of low-mass stars in the Galactic field \citep{andy08}. We are currently conducting a comprehensive census of magnetic activity in Praesepe to test directly the equivalence of the activity lifetime and spin-down timescale. For more on our activity analysis, see the presentation by Lemonias et al.\ elsewhere in these proceedings.  

\section{Current Status \& Future Plans}\label{future}
We recently began our second season of observations of Praesepe. Our observing strategy has changed somewhat, and the CCCP footprint now features less overlap and covers a total area of $\sim$40 deg$^2$. In parallel to this continued photometric monitoring, we are conducting a spectroscopic campaign to confirm the membership and measure the chromospheric activity of the \citet{adam2007} P50 stars. Our observations with the 2.4-m Hiltner Telescope at the MDM Observatory and using the Hydra multi-object spectrograph on the WIYN 3.5-m telescope, both on Kitt Peak, AZ, will enable us to confirm our targets as members based on their radial velocities, and to obtain H$\alpha$ line measurements to diagnose their levels of chromospheric activity. Moreover, we have proposed to obtain new X-ray imaging of Praesepe to expand the sample of cluster members with measured coronal activity, and to establish once and for all if Praesepe's X-ray luminosity function can be reconciled with that of the Hyades (see additional discussion in Lemonias et al., this volume).

Our next CCCP target, NGC 752, is a $\sim$1.1 Gyr cluster at a distance of 450 pc, and is the best studied, closest old cluster to the Sun. Stellar rotation at this age is generally not well constrained, and no rotation periods that we know of exist for NGC 752 members. The CCCP footprint for this cluster is comprised of two overlapping PTF fields such that the cluster core (which is roughly 0.5 deg across) is contained almost entirely on one chip in each field. Our observations began on 2010 Aug 22, and to date we have $>$300 individual observations of each field, so that the cluster core has been visited $>$600 times.  As the NGC 752 low-mass population is poorly defined, we are also beginning a spectroscopic campaign on this cluster this winter to determine the membership and activity level of candidate low-mass members in the cluster field.

\acknowledgments This research has made use of NASA's Astrophysics Data System Bibliographic Services, the SIMBAD database, operated at CDS, Strasbourg, France, the NASA/IPAC Extragalactic Database, operated by the Jet Propulsion Laboratory, California Institute of Technology, under contract with the National Aeronautics and Space Administration, and the VizieR database of astronomical catalogs \citep{Ochsenbein2000}. IRAF (Image Reduction and Analysis Facility) is distributed by the National Optical Astronomy Observatories, which are operated by the Association of Universities for Research in Astronomy, Inc., under cooperative agreement with the National Science Foundation. 
 
The Two Micron All Sky Survey was a joint project of the University of Massachusetts and the Infrared Processing and Analysis Center (California Institute of Technology). The University of Massachusetts was responsible for the overall management of the project, the observing facilities and the data acquisition. The Infrared Processing and Analysis Center was responsible for data processing, data distribution and data archiving.

\end{document}